\documentclass[lettersize,journal]{IEEEtran}
\usepackage{amsmath,amsfonts,amsthm}
\usepackage[ruled,commentsnumbered,linesnumbered]{algorithm2e}

\usepackage{array}
\usepackage{textcomp}
\usepackage[dvipsnames]{xcolor}
\usepackage{stfloats}
\ifCLASSOPTIONcompsoc
\usepackage[caption=false,font=normalsize,labelfont=sf,textfont=sf]{subfig}
\else
\usepackage[caption=false,font=footnotesize]{subfig}
\fi
\usepackage{url}
\usepackage{verbatim}
\usepackage{graphicx}
\usepackage{cite}
\usepackage[colorlinks,linkcolor=blue,citecolor=blue,urlcolor=blue]{hyperref} 
\usepackage{makecell}
\usepackage{multirow}
\usepackage{color}
\usepackage{booktabs}
\usepackage{amsmath}
\usepackage{colortbl}
\usepackage{bbm}
\usepackage{textcomp}
\usepackage{xcolor}
\usepackage{geometry}
\allowdisplaybreaks[4]
\usepackage{amssymb}
\usepackage{threeparttable}

\SetKwComment{Comment}{~$\triangleright$~}{}

\theoremstyle{remark}
\newtheorem{remark}{Remark}
\newtheorem{proposition}{Proposition}
\newtheorem{lemma}{Lemma}

\newtheorem{theorem}{Theorem}

\begin{document}

\title{Revisiting MUSIC: A Finite-Precision Perspective}
\author{Yiming Fang,~\IEEEmembership{Graduate Student Member,~IEEE}, 
    Li Chen,~\IEEEmembership{Senior Member,~IEEE},
    Ang Chen,
    and Weidong Wang

    \thanks{The authors are with the CAS Key Laboratory of Wireless-Optical Communications, University of Science and Technology of China, Hefei 230027, China (e-mail: fym1219@mail.ustc.edu.cn; chenli87@ustc.edu.cn; chenang1122@mail.ustc.edu.cn; wdwang@ustc.edu.cn).}\vspace{-0.7cm}

}



\maketitle
\begin{abstract}
The high computational complexity of the multiple signal classification (MUSIC) algorithm is mainly caused by the subspace decomposition and spectrum search, especially for frequent real-time applications or massive sensors. In this paper, we propose a low-complexity MUSIC algorithm from a finite-precision arithmetic perspective. First, we analyze the computational bottlenecks of the classic low-complexity randomized unitary-based MUSIC (RU-MUSIC), formulating this computational issue as an inner product problem. Then, a mixed-precision method is introduced to address this problem. Specifically, this method partitions summations in inner products into blocks, where intra-block computations use low-precision arithmetic and inter-block sums use high-precision arithmetic. To further improve computational accuracy, we develop an adaptive-precision method that supports adaptive block sizes and multiple precision levels. Finally, simulation results show that the proposed finite-precision MUSIC design achieves direction-of-arrival (DOA) estimation performance similar to that using full-precision arithmetic while reducing more than 50\% computational cost.
\end{abstract}


\section{Introduction}
\label{sec: intro}
\IEEEPARstart{D}{irection} of arrival (DOA) estimation for multiple signal sources is a critical challenge in radar, sensing, and communications \cite{8828025}. Among various DOA estimation methods, the multiple signal classification (MUSIC) algorithm is widely recognized for its superior resolution \cite{1143830}. Efficient implementation of the MUSIC algorithm is essential for frequent real-time applications or massive sensors.

Limited by the challenging implementation of the subspace decomposition and spectrum search, the classic MUSIC algorithm suffers from high computational complexity. Specifically, subspace decomposition is usually achieved by singular value decomposition (SVD), leading to its computational complexity of cubic order of the number of sensors. In spectrum search, a large number of search angles must be computed to achieve high resolution, which may cause much higher complexity than that of subspace decomposition \cite{4663945}.

To reduce the computational complexity of subspace decomposition, many methods with low-complexity randomized SVD (R-SVD) \cite{9325098} or avoiding SVD \cite{875461} have been proposed to reduce the computational complexity of subspace decomposition. Additionally, real-valued MUSIC variants, including unitary MUSIC (U-MUSIC) \cite{6704870} and real-valued MUSIC (RV-MUSIC) \cite{80927}, were proposed to reduce computational overhead by approximately 75\% compared to complex-valued implementations. Moreover, the authors in \cite{AHMAD2024155235} proposed to integrate the reduced-dimension MUSIC algorithm with the semi-definite programming (SDP) optimization solver to reduce the complexity of the MUSIC algorithm.

For spectrum search, the authors in \cite{1172124} proposed the root-MUSIC algorithm using polynomial rooting to avoid searching spectrum. Besides, the authors in \cite{6422415} introduced a compressed MUSIC (C-MUSIC) that only uses a limited search range to achieve spectrum search. Moreover, a randomized matrix sketching method was applied in \cite{9328487} to reduce the complexity of the spectrum search.


The aforementioned works on low-complexity MUSIC algorithms commonly implement the matrix computations using full-precision arithmetic. Recently, some works have explored reducing the computational complexity using finite-precision arithmetic in machine learning and wireless communications. Specifically, the authors in \cite{gupta2015deep} utilized finite-precision arithmetic to accelerate deep learning training. Besides, the authors in \cite{10845867} analyzed the impact of finite-precision arithmetic on communication performance and proposed a mixed-precision transceiver design. However, to the best of our knowledge, the realization of low computational complexity MUSIC algorithm taking advantage of finite-precision arithmetic is still an open problem.

Motivated by the above observations, in this paper, we propose a low-complexity MUSIC algorithm from a finite-precision perspective. First, the computational bottlenecks of the randomized U-MUSIC (RU-MUSIC) are analyzed, and then this computational issue is formulated as an inner product problem. Further, we introduce a mixed-precision (MP) method to address this problem. Specifically, this method partitions summations in inner products into blocks, where intra-block computations are implemented by low-precision arithmetic and inter-block sums are calculated by high-precision arithmetic. Moreover, an adaptive-precision (AP) method is presented to enhance computational accuracy, supporting adaptive block sizes and multiple precisions. Finally, simulation results demonstrate that the proposed finite-precision MUSIC design can achieve a close DOA estimation performance to that of full-precision arithmetic, while it can reduce over 50\% computational cost. 




\section{System Model}
\label{sec: sys}
\subsection{Signal Model}

We consider $N$ narrowband signals impinging on a uniform linear array (ULA) consisting of $M (M>N)$ sensors from $N$ different directions $\boldsymbol{\bar{\theta} }=\left[\bar{\theta}_1, \bar{\theta}_2, \cdots, \bar{\theta}_N \right]^\mathsf{T}$. The observation equation can be written as
\begin{align}\label{eq: signal_model}
    \mathbf{x}\left( t \right) & = \sum_{n=1}^N s_n(t) {\bf a}\left(\bar{\theta}_n\right) + {\bf n}(t) \notag\\
    &=\mathbf{A}\left( \boldsymbol{\bar{\theta} } \right) \mathbf{s}\left( t \right) +\mathbf{n}\left( t \right),~ t=1,2,\cdots, T,
\end{align}
where $\mathbf{A}\left( \boldsymbol{\bar{\theta} } \right) =\left[ \mathbf{a}\left( \bar{\theta} _1 \right) ,\cdots ,\mathbf{a}\left( \bar{\theta} _N \right) \right] \in \mathbb{C} ^{M\times N}$ is the steering matrix with the $n$-th column
\begin{align}
    \mathbf{a}\left( \bar{\theta}_n \right) =\left[ 1,e^{-\mathsf{j}\frac{2\pi}{\lambda}d\sin \bar{\theta}_n},\cdots ,e^{-\mathsf{j}\frac{2\pi}{\lambda}\left( M-1 \right) d\sin \bar{\theta}_n} \right] ^\mathsf{T}
\end{align}
denoting the steering vector of the $n$-th source with $\lambda$ representing the wavelength and $d = \lambda/2$ representing sensor spacing, $\mathsf{j}$ is imaginary unit, $\mathbf{s}\left( t \right) = \left[s_1(t), s_2(t), \cdots, s_N(t) \right]^\mathsf{T}\in \mathbb{C}^{N\times 1}$ is the incoherent signals at time instance $t$, ${\bf n}(t) \sim \mathcal{CN}({\bf 0},\sigma_n^2{\bf I}_{M})$ is the complex additive white Gaussian noise (AWGN), $\sigma_n^2$ is the variance of noise, and $T$ is the number of snapshots. Then, the covariance matrix of $\mathbf{x}\left( t \right)$ is 
\begin{align}\label{eq: em_covariance}
    {\bf R}_{\bf x} =\mathbb{E} \left\{ \mathbf{x}\left( t \right) \mathbf{x}^\mathsf{H}\left( t \right) \right\} \approx \frac{1}{T}\sum_{t=1}^{T}\mathbf{x}\left( t \right) \mathbf{x}^\mathsf{H}\left( t \right).
\end{align}

 
\vspace{-0.6cm}
\subsection{MUSIC and RU-MUSIC}
To estimate DOAs, i.e., $\boldsymbol{\bar{\theta} }$, a popular approach is the MUSIC algorithm, which contains the subspace decomposition and spectrum search given below \cite{1143830}.
\subsubsection{Subspace Decomposition} Using the covariance matrix ${\bf R}_{\bf x}$ in \eqref{eq: em_covariance}, we can compute its SVD as
\begin{align}
    \mathbf{R}_{\mathbf{x}}=\mathbf{U\Sigma U}^\mathsf{H}=\mathbf{U}_s\mathbf{\Sigma}_s\mathbf{U}_{s}^\mathsf{H}+\sigma_{n}^{2}\mathbf{U}_n\mathbf{U}_{n}^\mathsf{H},
\end{align}
where $\mathbf{U}_s \in \mathbb{C}^{M\times N}$ and $\mathbf{U}_n \in \mathbb{C}^{M\times (M-N)}$ is the signal subspace and noise subspace, respectively.
\subsubsection{Spectrum Search} Based on the signal subspace or noise subspace, for a set of discrete angles $\boldsymbol{\theta }=\left[\theta_1, \theta_2, \cdots, \theta_{F} \right]^\mathsf{T} \in \left[-90^\circ,90^\circ\right]$, the search function is given by
\begin{align}
    P\left( \theta_i \right) &=\frac{1}{\mathbf{a}^\mathsf{H}\left( \theta_i\right) \mathbf{U}_n\mathbf{U}_{n}^\mathsf{H}\mathbf{a}\left( \theta_i\right)} \label{eq: spectrum_n}\\
    &=\frac{1}{\mathbf{a}^\mathsf{H}\left( \theta_i \right) \left( \mathbf{I}_M-\mathbf{U}_s\mathbf{U}_{s}^\mathsf{H} \right) \mathbf{a}\left( \theta_i \right)},i=1,\cdots,F. \label{eq: spectrum_s}
\end{align}
For $N$ targets, $N$ peaks can be identified from \eqref{eq: spectrum_n} or \eqref{eq: spectrum_s}, enabling the estimation of the unknown DOAs.

Notably, the covariance matrix ${\bf R}_{\bf x}$ is complex-valued. Thus, the above classic MUSIC algorithm involves complex-valued computations, which poses challenges for real-time applications due to high computational complexity \cite{6704870}. To this end, RU-MUSIC was proposed in \cite{839978,9325098} to reduce the complex-valued computations to real-valued computations and alleviate the complexity of exact SVD. Based on unitary transformation, the real-valued covariance matrix is given by \cite{839978}
\begin{align} \label{eq: em_covariance_real}
    \mathbf{C}=\Re \left( \mathbf{Q}^\mathsf{H}\mathbf{R}_{\mathbf{x}}\mathbf{Q} \right) \in \mathbb{R}^{M\times M},
\end{align}
where the sparse matrices
\begin{align}
     \mathbf{Q}=\frac{1}{\sqrt{2}}\left[ \begin{matrix}
	\mathbf{I}&		\mathsf{j}\mathbf{I}\\
	\mathbf{J}&		-\mathsf{j}\mathbf{J}\\
\end{matrix} \right] ~\mathrm{or}~
\mathbf{Q}=\frac{1}{\sqrt{2}}\left[ \begin{matrix}
	\mathbf{I}&		\mathbf{0}&		\mathsf{j}\mathbf{I}\\
	\mathbf{0}^\mathsf{T}&		\sqrt{2}&		\mathbf{0}^\mathsf{T}\\
	\mathbf{J}&		\mathbf{0}&		-\mathsf{j}\mathbf{J}\\
\end{matrix} \right]
\end{align}
can be chosen for arrays with an even and odd number of sensors, $\mathbf{0}$ is zero vector and $\bf J$ is the exchange matrix with ones on its anti-diagonal and zeros elsewhere.

Then, based on R-SVD, we can obtain the rank-$K$ ($N<K<M$) SVD of $\mathbf{C}$ as
\begin{align}\label{eq: real_svd}
\mathbf{C}_K\approx\tilde{\mathbf{U}}_K\tilde{\mathbf{\Sigma}}_K\tilde{\mathbf{V}}_{K}^\mathsf{T},
\end{align}
where $\mathbf{C}_K$ is the best rank-$K$ approximation of $\bf C$, which approximates $\bf C$ by its first $K$ singular vectors and values, $\tilde{\mathbf{U}}_K \in \mathbb{R}^{M\times K}$, $\tilde{\mathbf{\Sigma}}_K\in \mathbb{R}^{K\times K}$, and $\tilde{\mathbf{V}}_{K}\in \mathbb{R}^{M\times K}$ are the right singular matrix, singular values matrix and left singular matrix of $\mathbf{C}_K$, respectively.

Further, given a set of angles $\boldsymbol{\theta }=\left[\theta_1, \theta_2, \cdots, \theta_{F} \right]^\mathsf{T}$, the search function can be expressed as \cite{80927}
\begin{align}
    P\left( \theta_i \right) =\frac{1}{\tilde{\mathbf{a}}^\mathsf{T}\left( \theta_i \right) \left( \mathbf{I}_M-\mathbf{\hat{E}}_s\mathbf{\hat{E}}_s^\mathsf{T} \right) \tilde{\mathbf{a}}\left( \theta_i \right)}, i = 1,\cdots,F, \label{eq: spectrum_s_real}
\end{align}
where $\mathbf{\hat{E}}_s = \tilde{\mathbf{U}}_K(:,1:N)$ is the approximate signal subspace, and
\begin{align*}
    &\tilde{\mathbf{a}}\left( \theta \right) =\sqrt{2}\big[  \cos \left( \frac{\pi}{\lambda}\left( M-1 \right) d\sin \theta \right), \cdots, 
    \cos \left( \frac{\pi}{\lambda}d\sin \theta \right), \notag \\
    & -\sin \left( \frac{\pi}{\lambda}d\sin \theta \right), \cdots, 
    -\sin \left( \frac{\pi}{\lambda}\left( M-1 \right) d\sin \theta \right) \big]^\mathsf{T} \in \mathbb{R}^{M}.
\end{align*}



\begin{algorithm}[t]
\label{alg: RU-MUSIC}
    \DontPrintSemicolon
    \SetAlgoNlRelativeSize{-1}
    \caption{RU-MUSIC}
    \KwIn{${\bf C} \in \mathbb{R}^{M\times M}$, an integer $K$ with $N<K<M$, and a set of angles $\boldsymbol{\theta }=\left[\theta_1, \theta_2, \cdots, \theta_{F} \right]^\mathsf{T}$}
    \KwOut{Estimated DOAs.}

    \tcp{R-SVD (subspace decomposition)}
    
    Initialize a random Gaussian matrix ${\bf \Omega}\in\mathbb{R} ^{M\times K}$ \;

    $\mathbf{B}=\mathbf{C\Omega }$ \Comment*[r]{$\mathcal{O}(M^2K)$}

    Compute $\mathbf{T} \in \mathbb{R}^{M\times K}$ via economy size QR decomposition: $\left[ \mathbf{T}, \sim \right] =\mathtt{qr}\left( \mathbf{B},0 \right)$ \Comment*[r]{$\mathcal{O}(MK^2)$}

    $\mathbf{D}=\mathbf{T}^\mathsf{T}\mathbf{C}$ \Comment*[r]{$\mathcal{O}(M^2K)$}

    Compute the economy size SVD: $\hat{\mathbf{U}}\tilde{\mathbf{\Sigma}}_K\tilde{\mathbf{V}}_{K}^\mathsf{T}=\mathtt{svd}\left( \mathbf{D},0 \right)$ \Comment*[r]{$\mathcal{O}(MK^2)$}

    $\tilde{\mathbf{U}}_K=\mathbf{T}\hat{\mathbf{U}}$ \Comment*[r]{$\mathcal{O}(MK^2)$}

    \tcp{Spectrum Search} 

    Assign the signal subspace $\mathbf{\hat{E}} = \tilde{\mathbf{U}}_K(:,1:N)$. Given $F$ discrete angles $\{\theta_i \}_{i=1}^F$ compute the spectrum $P\left( \theta_i \right) =\frac{1}{\tilde{\mathbf{a}}^\mathsf{T}\left( \theta_i \right) \left( \mathbf{I}_M-\mathbf{\hat{E}}_s\mathbf{\hat{E}}_s^\mathsf{T} \right) \tilde{\mathbf{a}}\left( \theta_i \right)}$ \Comment*[r]{$\mathcal{O}(MNF)$}

    Estimate $N$ DOAs by identifying the peaks of $P\left( \theta_i \right)$ \;
    
\end{algorithm}

Note that \eqref{eq: em_covariance_real} can be an efficient computation due to the sparse structure of $\mathbf{Q}$ \cite{80927,382406}, which can be negligible compared to the computations of \eqref{eq: real_svd} and \eqref{eq: spectrum_s_real}. Moreover, the RU-MUSIC algorithm is summarized in \textit{Algorithm} \ref{alg: RU-MUSIC}. On the one hand, since the RU-MUSIC algorithm only requires real-valued computations, the RU-MUSIC algorithm can reduce about 75\% computational burdens compared to the classic MUSIC algorithm. On the other hand, the RU-MUSIC algorithm uses R-SVD to obtain the rank-$K$ ($N<K<M$) approximation, whose computational complexity is $\mathcal{O}(M^2K)$, much lower than that of the exact SVD in the MUSIC algorithm.



\begin{remark}[\textit{Challenges of Implementing Low-Complexity MUSIC}]
    Even for the RU-MUSIC algorithm, its implementation remains challenging due to two key computational bottlenecks. First, in subspace decomposition, \textit{Steps 3 and 5} of \textit{Algorithm} \ref{alg: RU-MUSIC} incur a computational complexity of $\mathcal{O}(M^2K)$, which dominants and remains the main computational bottleneck. Second, during spectrum search, the computational complexity of computing the spectrum in \textit{Step 9} is $\mathcal{O}(MNF)$. To achieve high-resolution angle estimation, $F$ must be set to a large value (e.g. $F=1000$ with resolution degree $\Delta \theta = 0.18^\circ$), leading to substantial computational overhead. \textit{To alleviate these bottlenecks, we propose replacing full-precision arithmetic ($\mathtt{fp64}$) with finite-precision arithmetic in the aforementioned corresponding steps (i.e. Steps 3, 5, and 9) of \textit{Algorithm} \ref{alg: RU-MUSIC}.}
\end{remark}

\section{Finite-Precision MUSIC Design}
In this section, we propose a finite-precision MUSIC design. First, we formulate the above computational problem as one inner
product procedure and present an MP method. Then, to further improve the accuracy of inner products, the AP method is introduced.

\subsection{Mixed-Precision Method}

\begin{algorithm}[t]
    \DontPrintSemicolon
    \SetAlgoNlRelativeSize{-1}
    \label{alg: MP}
    \caption{Mixed-Precision Method with precision $u_l$ and $u_h$}
    \KwIn{${\bf b,c} \in \mathbb{R}^{M\times 1}$, and block size $B$}
    \KwOut{$y = \mathbf{b}^{\mathsf{T}}\mathbf{c}$}

    $p = \lceil M/B \rceil$
    
    \For{$k = 1 : p$}{
    $y_k = 0$, $\tau_{b} = 1+(k-1)B$, $\tau_{e} = \min (kB,M)$

    \For{$i= \tau_{b}:\tau_{e}$}{
     $y_k = y_k + b_i c_i$ in precision $u_l$ 
    }
    }
    $y = \sum_{k=1}^{p} y_k$ in precision $u_h$ 
    
\end{algorithm}

The computational bottleneck in \textit{Algorithm} \ref{alg: RU-MUSIC} arises from matrix-matrix products, which consist of the sequential inner products. Thus, to reduce the computational complexity, it is essentially to accelerate the calculation of the inner product, which can be implemented through finite-precision arithmetic. 



Notably, directly applying finite-precision arithmetic, especially low-precision arithmetic, for inner products introduces substantial computational errors \cite{higham2022mixed}. To mitigate these errors, MP methods have been investigated in numerical linear algebra \cite{FAM} and signal processing \cite{10845867}, which is summarized in \textit{Algorithm} \ref{alg: MP}. Specifically, this method involves partitioning summations in inner products into blocks with block size $B$, where intra-block computations use low precision $u_l$ and then inter-block sums in high precision $u_h$. Here, we use it for efficient implementation of the MUSIC algorithm. Nevertheless, the MP method is restricted to artificially fixed block sizes and two precision levels, and cannot support adaptive block sizes or multiple precisions. To this end, we present a generalized AP method in the following subsection.

\begin{remark}[\textit{Complexity Analysis of Algorithm \ref{alg: MP}}]
    We define a ratio of $q_l:q_h$ for the costs of low-precision and high-precision arithmetic. For example, we have a ratio of $1:2:4$ for the costs of $\mathtt{fp16}$, $\mathtt{fp32}$, and $\mathtt{fp64}$ arithmetic \cite{FAM}. Then given the ratio $q_l$ and $q_h$, the number of additions is $\mathcal{C}_{\rm add}^{\rm MP} = p q_l(B - 1) + q_h(p - 1)$, and the number of multiplications is $\mathcal{C}_{\rm multi}^{\rm MP} =  q_lB$.
\end{remark}

\subsection{Adaptive-Precision Method}
First, following \cite{higham2022mixed} and \cite{doi:10.1137/22M1522619}, we define $u$ as the unit roundoff of the precision used and introduce the AP method shown in \textit{Algorithm} \ref{alg: AP}. This method calculates the inner product $y = \mathbf{b}^{\mathsf{T}}\mathbf{c}$ using $p(p\geq 2)$ precision levels $u_1 < u_2< \cdots < u_p$. Vectors $\bf b$ and $\bf c$ are partitioned into $p$ groups based on $\{\mathcal{G}_k\}_{k=1}^p$, and the partial inner products $y_k = \sum_{i \in \mathcal{G}_k}b_i c_i $ corresponding to group $\mathcal{G}_k$ is computed in precision $u_k$. All the partial inner products are then summed using the highest precision $u_1$. 

\begin{algorithm}[t]
    \DontPrintSemicolon
    \SetAlgoNlRelativeSize{-1}
    \label{alg: AP}
    \caption{Adaptive-Precision Method in $p$ precision levels $u_1 < u_2< \cdots < u_p$}
    \KwIn{${\bf b,c} \in \mathbb{R}^{M\times 1}$, $i = 1:M$, and groups $\{\mathcal{G}_k\}_{k=1}^p$}
    \KwOut{$y = \mathbf{b}^{\mathsf{T}}\mathbf{c}$}

    \For{$k = 1 : p$}{
    $y_k = 0$
    
    \For{$i\in \mathcal{G}_k$}{
     $y_k = y_k + b_i c_i$ in precision $u_k$ 
    }
    }
    $y = \sum_{k=1}^p y_k$ in precision $u_1$ 
    
\end{algorithm}

Since \textit{Algorithm} \ref{alg: AP} has been well developed, we next establish the form of groups $\{\mathcal{G}_k\}_{k=1}^p$ through rounding error analysis. Before conducting this analysis, we present the following useful lemma.
\begin{lemma}[\textit{Inner products}{\cite[Theorem 4.2]{doi:10.1137/120894488}}]
\label{lem: uniform}
    Let $y= \mathbf{b}^{\mathsf{T}}\mathbf{c}$, where ${\bf b,c} \in \mathbb{R}^{M\times 1}$, be evaluated in the finite-precision arithmetic with uniform precision $u$. Provided that no underflow or overflow is encountered, the computed $\hat{y}$ satisfies
    \begin{align}
        \left| \hat{y} - y\right| \leq Mu \left| \mathbf{b}\right|^\mathsf{T}\left| \mathbf{c}\right|.
    \end{align}
\end{lemma}

Then, we can provide the rounding error analysis of \textit{Algorithm} \ref{alg: AP} in the following theorem.
\begin{theorem}[\textit{Rounding error analysis of Algorithm \ref{alg: AP}}] \label{the: bwd}
    Let ${\bf b,c} \in \mathbb{R}^{M\times 1}$, and let $y = \mathbf{b}^{\mathsf{T}}\mathbf{c}$ be computed using \textit{Algorithm} \ref{alg: AP}. Then the rounding error $\gamma_{\rm AP}$ satisfies
    \begin{align} \label{eq: bwd}
        \gamma_{\rm AP} \le \varepsilon +\left( 1+\varepsilon \right) \sum_{k=1}^p{m_ku_k\left( 1+u_k \right) ^2\beta _k}, 
    \end{align}
    where $m_k$ is the size of group $\mathcal{G} _k$, and
    \begin{align} \label{eq: bwd_para}
        \varepsilon =\left( p-1 \right) u_1,~ \gamma_{\rm AP} =\frac{\left| \hat{y}-y \right|}{\left| \mathbf{b} \right|^{\mathsf{T}}\left| \mathbf{c} \right|}, ~\beta_k=\frac{\sum_{i\in \mathcal{G} _k}{\left| b_ic_i \right|}}{\left| \mathbf{b} \right|^{\mathsf{T}}\left| \mathbf{c} \right|}.
    \end{align}
\end{theorem}
\begin{IEEEproof}
    Using \textit{Lemma} \ref{lem: uniform}, the partial inner products $\hat{y}_k$ satisfies
    \begin{align} \label{eq: 1}
        \left| \hat{y}_k-y_k \right|\le m_ku_k\left( 1+u_k \right) ^2\sum_{i\in \mathcal{G} _k}{\left| b_ic_i \right|}\triangleq \xi _k,
    \end{align}
    where the term $\left( 1+u_k \right) ^2$ is caused by converting $b_i$ and $c_i$ to precision $u_k$ {\cite[Sec. 2.2]{higham2002accuracy}}. Further, we define $\tilde{y}=\sum_{i=1}^p \hat{y}_k$ as the \textit{exact} sum of the partial inner products $\hat{y}_k$ and have
    \begin{align} \label{eq: 2}
        \left| \tilde{y}-y \right|\le \sum_{k=1}^p{\xi _k},
    \end{align}
    and the computed $\hat{y}$ satisfies
    \begin{align}
        \left| \hat{y}-\tilde{y} \right|\overset{(a)}{\le} \varepsilon \sum_{k=1}^p{\left| \hat{y}_k \right|}
        \overset{\eqref{eq: 1}}{\le}\varepsilon \sum_{k=1}^p{\left( \sum_{i\in \mathcal{G} _k}{\left| b_ic_i \right| + \xi _k} \right)}, \label{eq: 3}
    \end{align}
    where $(a)$ uses the results in \cite[Sec. 4.2]{higham2002accuracy}, $\varepsilon =\left( p-1 \right) u_1$, and converting $\hat{y}_k$ to precision $u_1$ does not introduce any error due to $u_1 < u_k$. Combining \eqref{eq: 2} and \eqref{eq: 3}, we can obtain
    \begin{align}
        \left| \hat{y}-y \right|&\le \left| \hat{y}-\tilde{y} \right|+\left| \tilde{y}-y \right| \\
        &\overset{(b)}{\le} \varepsilon \left| \mathbf{b} \right|^{\mathsf{T}}\left| \mathbf{c} \right|+\left( 1+\varepsilon \right) \sum_{k=1}^p{\xi _k}, \label{eq: 4}
    \end{align}
    where $\sum_{k=1}^p{\sum_{i\in \mathcal{G} _k}{\left| b_ic_i \right|}}=\sum_{i=1}^M{\left| b_ic_i \right|}=\left| \mathbf{b} \right|^{\mathsf{T}}\left| \mathbf{c} \right|$ at $(b)$. Finally, dividing $\left| \mathbf{b} \right|^{\mathsf{T}}\left| \mathbf{c} \right|$ at both side in \eqref{eq: 4}, we can obtain the rounding error in \eqref{eq: bwd}. And the proof ends.
\end{IEEEproof}

\textit{Theorem} \ref{the: bwd} reveals that the ratio $\beta_k$ in \eqref{eq: bwd_para} controls the value of the rounding error when other parameters are fixed. Furthermore, the ratio $\beta_k$ quantifies the relative size of elements in group $\mathcal{G}_k$ compared to all elements in the inner products. Consequently, \textit{Theorem} \ref{the: bwd} implies that assigning elements with smaller magnitudes to lower-precision groups minimizes rounding error. Based on this analysis and the ratio $\beta_k$, we can obtain the specific form groups $\{\mathcal{G}_k\}_{k=1}^p$.
\begin{proposition}[\textit{The specific form of groups $\{\mathcal{G}_k\}_{k=1}^p$}]\label{prop: ap}
    Given a target accuracy $\gamma (\gamma \geq u_1)$ and $p(p\geq 2)$ groups, the groups $\{\mathcal{G}_k\}_{k=1}^p$ in \textit{Algorithm} \ref{alg: AP} are defined as
    \begin{align}\label{eq: form_G}
        \mathcal{G} _k=\left\{ i \mid \left| b_ic_i \right|\in \mathcal{I} _k, i=1:M \right\},~k=1:p,
    \end{align}
    where the intervals $\mathcal{I}_k$ can be expressed as
    \begin{align*}
        \mathcal{I} _k=\begin{cases}
	\left( \gamma \left| \mathbf{b} \right|^{\mathsf{T}}\left| \mathbf{c} \right|/u_2,+\infty \right),&		k=1\\
	\left( \gamma \left| \mathbf{b} \right|^{\mathsf{T}}\left| \mathbf{c} \right|/u_{k+1},\gamma \left| \mathbf{b} \right|^{\mathsf{T}}\left| \mathbf{c} \right|/u_k \right],&		k=2:p-1\\
	\left[ 0,\gamma \left| \mathbf{b} \right|^{\mathsf{T}}\left| \mathbf{c} \right|/u_p \right],&		k=p\\
    \end{cases}.
    \end{align*}
    Then, based on \eqref{eq: form_G}, the rounding error $\gamma_{\rm AP}$ satisfies
    \begin{align} \label{eq: bwd_gamma}
        \gamma_{\rm AP} \le \varepsilon + c\gamma \sim \mathcal{O}(\gamma),
    \end{align}
    where $c = (1+\varepsilon)\sum_{k=1}^{p}m_k^2(1+u_k)^2$.
\end{proposition}
\begin{IEEEproof}
    Note that $\beta_k \leq m_k \gamma /u_k, k=1:p$ based on \eqref{eq: form_G}. Thus, from \eqref{eq: bwd}, we can obtain \eqref{eq: bwd_gamma}.
\end{IEEEproof}
\textit{Proposition} \ref{prop: ap} demonstrates that, for adaptive-precision inner products, elements with larger magnitudes should be maintained at higher precision, while those with smaller magnitudes can be assigned lower precisions. Additionally, given a target accuracy $\gamma$, we can ensure the rounding error in \textit{Algorithm} \ref{alg: AP} is at most in $\mathcal{O}(\gamma)$ based on groups $\{\mathcal{G}_k\}_{k=1}^p$ in \eqref{eq: form_G}.

Finally, the AP method is developed by combining \textit{Algorithm} \ref{alg: AP} and \textit{Proposition} \ref{prop: ap}. Furthermore, we can use the AP method to implement the computational bottleneck part (i.e. \textit{Steps 3, 5, and 9}) in \textit{Algorithm} \ref{alg: RU-MUSIC}.

\begin{remark}[\textit{Complexity Analysis of Algorithm \ref{alg: AP}}]
    Given a ratio $q_k$ of precision $u_k$ corresponding to group $\mathcal{G}_k$ for $k=1:p$, the number of additions is $\mathcal{C}_{\rm add}^{\rm AP} = \sum_{k=1}^p q_k(m_k - 1) + q_1(p - 1)$, and the number of multiplications is $\mathcal{C}_{\rm multi}^{\rm AP} = \sum_{k=1}^p q_km_k$.
\end{remark}


\section{Simulation Results and Discussions}
\label{sec: sim}

\begin{figure*}[t]
    \centering
    \subfloat[]{\includegraphics[width=0.3\textwidth]{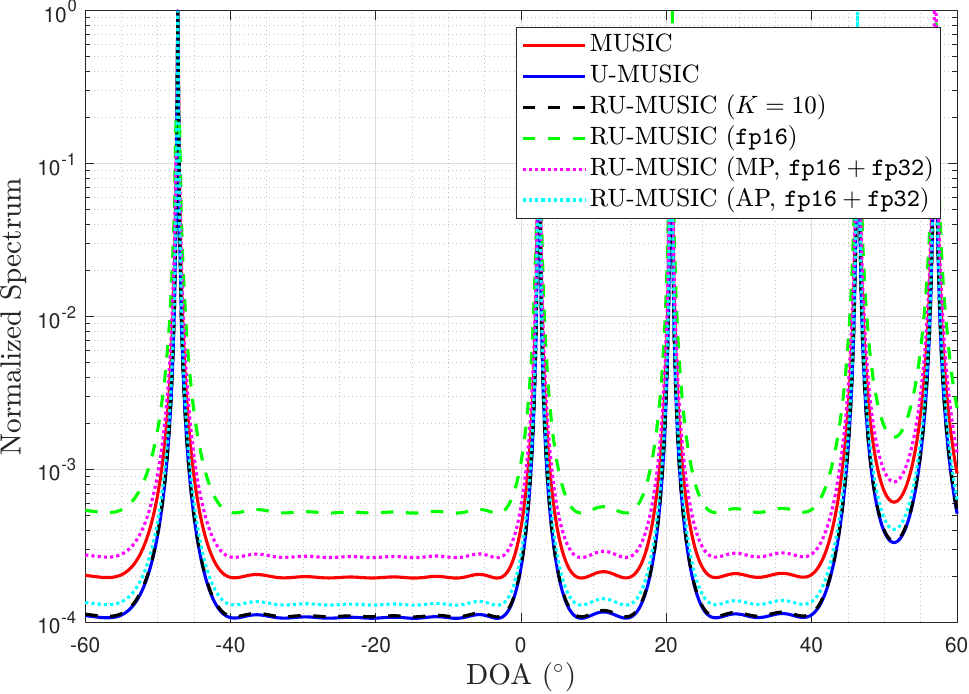}
    \label{fig: demo}}
    \subfloat[]{\includegraphics[width=0.31\textwidth]{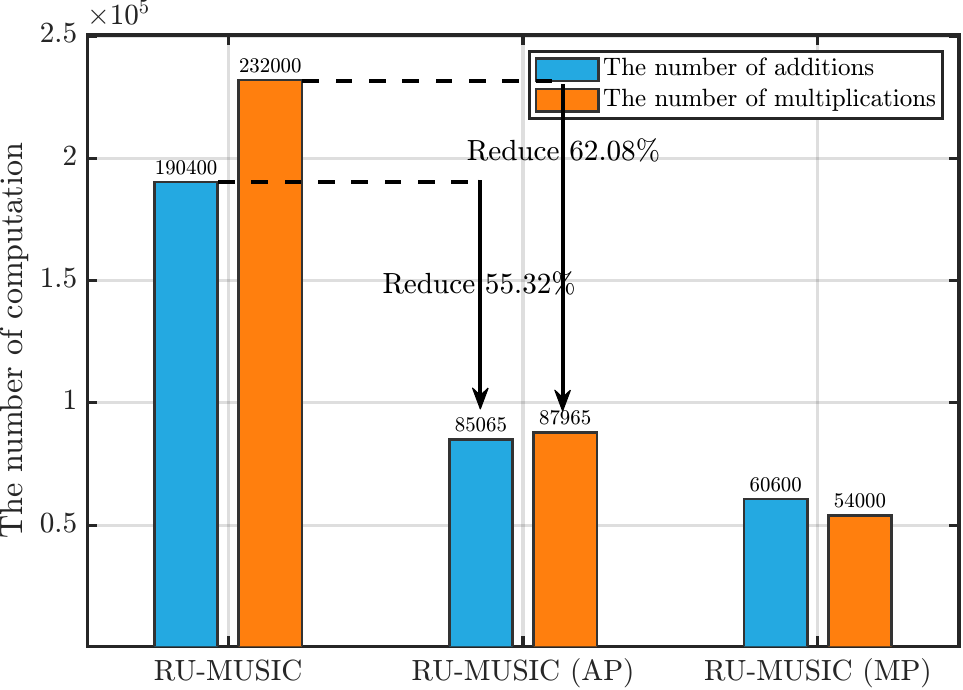}
    \label{fig: CC}}
    \subfloat[]{\includegraphics[width=0.3\textwidth]{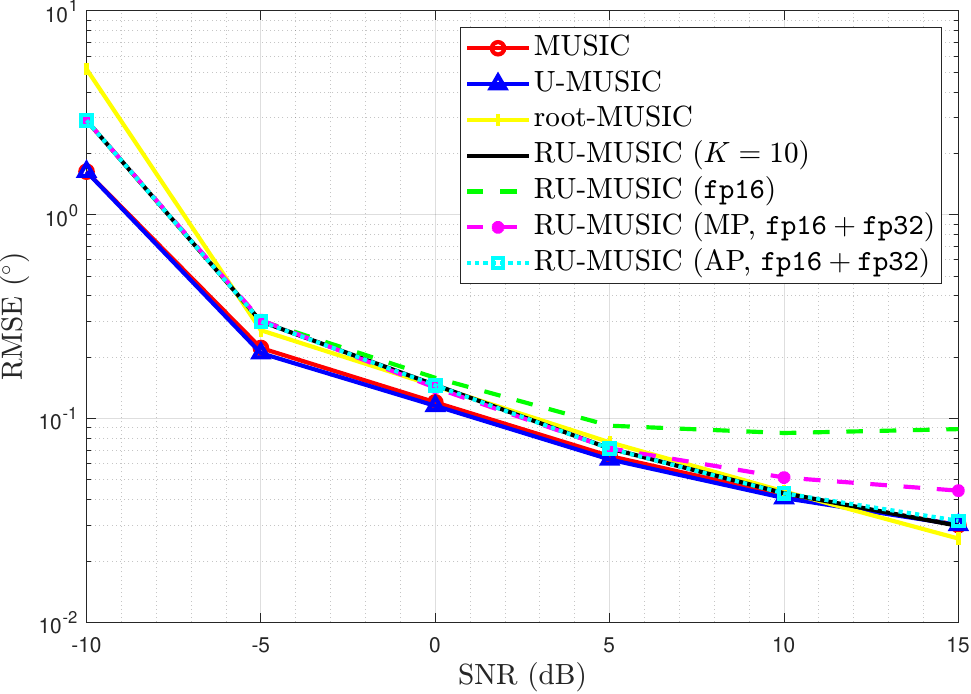}
    \label{fig: RMSE}}
    \caption{Performance and computational cost. (a) Estimated spectrum of various algorithms in a single trial with SNR = 20 dB. (b) The number of additions and multiplications of various algorithms. (c) RMSE of various algorithms.}
    \label{fig: sim}
\end{figure*}

In this section, we will provide numerical results to verify the performance of the proposed finite-precision MUSIC design. The simulation parameters are provided as follows. 
\subsubsection{Parameters of Finite-Precision Arithmetic} The authors in \cite{higham2019simulating} provided a function that can be utilized to simulate $\mathtt{fp32}$ with $u=2^{-24}$, $\mathtt{fp16}$ with $u=2^{-11}$, and other low-precision arithmetic. For the analysis of computational cost, we consider a ratio of $1:2:4$ for the costs of $\mathtt{fp16}$, $\mathtt{fp32}$, and $\mathtt{fp64}$ arithmetic \cite{FAM}.
\subsubsection{Parameters of DOA estimation} Similar to \cite{9991096}, we assume that each DOA follows a uniform distribution $\mathcal{U}[-60^\circ, 60^\circ]$ and $N$ DOAs in each trial are sampled with at least $10^\circ$ separation to keep sources resolvable. We set $\mathcal{L}=1000$ Monte-Carlo trials for each signal-to-noise ratio (SNR) level to obtain the root mean square error (RMSE) as
\begin{align}
    \mathrm{RMSE}=\sqrt{\frac{1}{\mathcal{L} N}\sum_{\ell =1}^{\mathcal{L}}{\sum_{n=1}^N{\left( \hat{\theta}_{\ell ,n}-\theta _{\ell ,n} \right) ^2}}}, \notag
\end{align}
where $\hat{\theta}_{\ell ,n} $ is the estimate of the $k$-th DOA $\theta _{\ell ,n}$ in the $\ell$-th Monte-Carlo trial. The number of DOAs, sensors, and snapshots is $N=5$, $M=20$, and $T=40$, respectively. Additionally, in R-SVD, we set $K=10$, and the number of discrete angles is $F=1500$ for spectrum search. For the MP method, the block size is $B = 2$. For the AP method, the target accuracy is $\gamma = 2^{-16}$.

As shown in Fig. \ref{fig: sim}, we evaluate the performance and computational cost of seven algorithms: MUSIC, U-MUSIC, root-MUSIC, RU-MUSIC, RU-MUSIC with pure low-precision, MP-based RU-MUSIC, and AP-based RU-MUSIC. First, Fig. \ref{fig: demo} illustrates the estimated spectrum of these algorithms in a single trial at SNR = 20 dB. Notably, the AP-based RU-MUSIC achieves a spectral approximation that closely aligns with U-MUSIC and RU-MUSIC, outperforming both the MP-based variant and pure low-precision implementations. Then, to quantify the computational cost, Fig. \ref{fig: CC} compares the number of additions and multiplications across different methods. Specifically, compared with full-precision RU-MUSIC ($\mathtt{fp64}$), the AP-based method reduces additions and multiplications by $55.32\%$ and $62.08\%$, respectively. At the same time, it incurs only marginal overhead compared to the MP method. Finally, Fig. \ref{fig: RMSE} analyzes RMSE performance. At low SNR levels, low-precision arithmetic ($\mathtt{fp16}$) achieves accuracy comparable to full-precision baselines. At high SNR levels, AP-based RU-MUSIC maintains near full-precision RMSE performance, whereas low-precision and MP-based implementations exhibit significant degradation. This confirms the AP method’s superior performance against rounding errors. Collectively, AP-based RU-MUSIC offers a favorable balance between computational efficiency and estimation accuracy. When computing resources are severely limited, MP-based RU-MUSIC is a good choice.

\section{Conclusions}
\label{sec: con}
In this paper, we have proposed a finite-precision MUSIC design for DOA estimation. First, we have analyzed the computational bottlenecks of the RU-MUSIC algorithm and formulated this computational issue as an inner product problem. Then, the MP method has been presented to address this problem. To further enhance computational accuracy, we have introduced an AP method that supports adaptive block sizes and multiple precisions. Finally, simulation results have revealed that the proposed finite-precision MUSIC design can achieve a close estimation performance to that using full-precision arithmetic, while it can reduce the number of additions and multiplications by $55.32\%$ and $62.08\%$, respectively.

\bibliographystyle{IEEEtran}
\bibliography{reference}

\end{document}